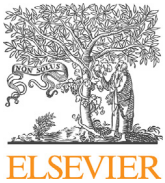
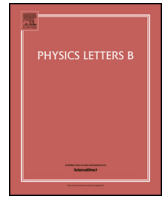

# One generation of standard model Weyl representations as a single copy of $\mathbb{R}\otimes\mathbb{C}\otimes\mathbb{H}\otimes\mathbb{O}$

N. Furey [a,b,*], M.J. Hughes [c]

[a] *Iris Adlershof, Humboldt-Universität zu Berlin, Zum Grossen Windkanal 2, Berlin, 12489, Germany*
[b] *AIMS South Africa, 6 Melrose Road, Muizenberg, Cape Town, 7945, South Africa*
[c] *Imperial College London, Prince Consort Rd, Kensington, London, SW7 2BW, United Kingdom of Great Britain and Northern Ireland*

A B S T R A C T

Peering in from the outside, $\mathbb{A}:=\mathbb{R}\otimes\mathbb{C}\otimes\mathbb{H}\otimes\mathbb{O}$ looks to be an ideal mathematical structure for particle physics. It is 32 $\mathbb{C}$-dimensional: exactly the size of one full generation of fermions. Furthermore, as alluded to earlier in [1], it supplies a richer algebraic structure, which can be used, for example, to replace $SU(5)$ with the $SU(3)\times SU(2)\times U(1)/\mathbb{Z}_6$ symmetry of the standard model.

However, this line of research has been weighted down by a difficulty known as the fermion doubling problem. That is, a satisfactory description of $SL(2,\mathbb{C})$ symmetries has so far only been achieved by taking two copies of the algebra, instead of one. Arguably, this doubling of states betrays much of $\mathbb{A}$'s original appeal. In this article, we solve the fermion doubling problem in the context of $\mathbb{A}$.

Furthermore, we give an explicit description of the standard model symmetries, $\breve{g}_{sm}$, its gauge bosons, Higgs, and a generation of fermions, each in the compact language of this 32 $\mathbb{C}$-dimensional algebra. Most importantly, we seek out the subalgebra of $\breve{g}_{sm}$ which is invariant under the complex conjugate - and find that it is given by $su(3)_C \oplus u(1)_{EM}$. Could this new result provide a clue as to why the standard model symmetries break in the way that they do?

## 1. Introduction

### 1.1. Why $\mathbb{R}\otimes\mathbb{C}\otimes\mathbb{H}\otimes\mathbb{O}$?

With the need of particle physics to adapt to frequent experimental update, it has made sense over the years to describe it with the most accommodating mathematics. The $n\times n$ matrices, or more generally, tensor algebras, are easily tailored to suit any theory. And with no doubt, it is precisely this flexibility which has enabled our models to weather constant changes over the years.

However, now that the experimental dust seems to have largely settled, it may be time to rethink our strategy. Could we learn more about particle physics by pairing it to a framework which has a unique structure of its own?

Oddly enough, there do exist mathematical objects which independently display some of the standard model's features. Here we consider in depth the Dixon algebra $\mathbb{A}:=\mathbb{R}\otimes\mathbb{C}\otimes\mathbb{H}\otimes\mathbb{O}$, the tensor product over $\mathbb{R}$ of the only four finite-dimensional normed division algebras over the reals.

*So, what features exactly*, one might ask, *are lacking in our current framework, but are automatic in* $\mathbb{A}$?

(1) $\mathbb{A}$ is 32$\mathbb{C}$-dimensional. This is precisely what is needed to encode the degrees of freedom of one generation of standard model fermions, including a right-handed neutrino. (Explicitly, we are referring to the number of off-shell degrees of freedom of the standard model's fermion fields.)

(2) $\mathbb{A}$'s combined left- and right-multiplication algebras are isomorphic to the Clifford algebra $\mathbb{C}l(10)$, and so one has ready access to $Spin(10)$ and its descendants: Pati-Salam, Georgi-Glashow, LR symmetric, and the standard model.

In fact, we will look to have $\mathbb{A}$ transform as a complex (**16**; **2**, **1**) left-handed Weyl representation under $spin(10)\oplus spin(1,3)$ in an upcoming paper, [2]. For the purposes of this current article, we will use this $spin(10)$ representation merely as a stepping stone to the standard model. (Also, we point out that this model will need to be augmented in future work so as to describe the infinite-dimensional unitary representations of spacetime symmetries.)

(3) However, unlike $\mathbb{C}l(10)$, $\mathbb{A}$ and its multiplication algebras have a richer inner structure. This is thanks to the division alge-

* Corresponding author.
  *E-mail addresses:* furey@physik.hu-berlin.de (N. Furey), mia.j.hughes@gmail.com (M.J. Hughes).

https://doi.org/10.1016/j.physletb.2022.136959
0370-2693/© 2022 The Author(s). Published by Elsevier B.V. This is an open access article under the CC BY license (http://creativecommons.org/licenses/by/4.0/). Funded by SCOAP³.

bras that comprise them. So whereas popular GUT models employ *ad hoc* Higgs fields and vacua to guarantee that gauge symmetries break in the right way, the algebra $\mathbb{A}$ may instead provide some rational guidance. Whether $\mathbb{A}$'s extra algebraic structure leads to the exclusion of larger symmetry groups, or merely their breaking at low energies, is yet to be seen.

For the standard model, what one needs is some mathematical object which justifies the *separate* factors of $SU(3)_C$, $SU(2)_L$, and $U(1)_Y$, in addition to the *separate* spacetime symmetries. However, pinning a construction down which can accomplish something along these lines is more difficult than one might imagine. The devil is in the detail.

*1.2. The fermion doubling problem*

At first glance, division algebraic particle physics may look to be a sure bet. After all, octonions readily describe $SU(3)$, the quaternions beget $SU(2)$, and the complex numbers enable $U(1)$. Indeed, a long history of proposals, [3–29], shows no lack of enthusiasm for a possibility that seems to be within reach.

So what then is the catch?

In short, the recurring challenges facing division algebraic endeavours have largely come down to peculiarities not in the standard model's gauge group, but rather in its *representations*:

(I) The standard model's electroweak sector is chiral.
(II) The standard model has three generations.
(III) The standard model's most efficient description does not include a linearly independent account of its redundant antiparticle fields.

In a recent paper, [1], it was explained how it is possible to arrive at a chiral theory by identifying the model's gauge symmetry with the unitary symmetries of division algebraic ladder operators. In the context of unbroken gauge symmetries, three generation structure has been identified in [13], [20]. However, in these recent papers, we tend to find a seemingly unnecessary doubling of fermionic states. This is a problematic state of affairs if our ultimate aim is to eventually match these representations to the quantum fields of the standard model.

In no place is this awkwardness more punitive than in the context of an $\mathbb{R} \otimes \mathbb{C} \otimes \mathbb{H} \otimes \mathbb{O}$ model. Under the typically assumed left action on $\mathbb{A}$ by the bivectors of its $\mathbb{C}l(10)$ multiplication algebra, we find $\mathbb{A}$ transforming under $Spin(10)$ as a $\mathbf{16} \oplus \mathbf{16}^*$.

Although seemingly innocuous, this is in fact a problematic state of affairs. For our current purposes, we would like to use $\mathbb{A}$ to describe a set of $2\mathbb{C}$ dimensional off-shell left-handed Weyl representations of $SL(2,\mathbb{C})$. In other words, we require *two copies* of $Spin(10)$'s $\mathbf{16}$, which are then linked to each other via $SL(2,\mathbb{C})$ transformations. Traditionally, in order to acquire a second $\mathbf{16}$, authors have opted to double the number of degrees of freedom from $\mathbf{16} \oplus \mathbf{16}^*$ to $\mathbf{16} \oplus \mathbf{16} \oplus \mathbf{16}^* \oplus \mathbf{16}^*$. In other words, *one generation is then no longer identified with one copy of $\mathbb{R} \otimes \mathbb{C} \otimes \mathbb{H} \otimes \mathbb{O}$*.

But this simple identification

$$\mathbb{R} \otimes \mathbb{C} \otimes \mathbb{H} \otimes \mathbb{O} \quad \leftrightarrow \quad \text{one generation}$$

was vital to $\mathbb{A}$'s original appeal.

What we demonstrate now is that it is indeed possible to recover what was originally envisioned: the one-to-one correspondence between $\mathbb{A}$ and one off-shell generation.

Our solution may easily be translated into the language of matrices and column vectors, where it resolves the analogous fermion doubling problem for $\mathbb{C}l(10)$ models involving the usual $32\mathbb{C}$-dimensional spinor. For literature on alternate fermion doubling problems, please see [30–33].

## 2. Solution to the fermion doubling problem

*2.1. The algebra $\mathbb{A}$*

Before explaining our resolution of the fermion doubling problem, we will need to equip the reader with some practical tools.

Let us define a complex quaternion as

$$c_\mu \epsilon_\mu \qquad \text{for} \quad \mu = 0 \ldots 3, \tag{1}$$

where $c_\mu \in \mathbb{C}$, $\epsilon_0 = 1$, and $\epsilon_i \epsilon_j = -\delta_{ij} + \varepsilon_{ijk} \epsilon_k$ for $i, j, k \in \{1, 2, 3\}$. Of course, $\varepsilon_{ijk}$ is the usual totally anti-symmetric tensor with $\varepsilon_{123} = 1$. The complex imaginary unit will be denoted as $i$, and is seen to commute with the $\epsilon_\mu$.

Similarly, we define a complex octonion as

$$c'_\nu e_\nu \qquad \text{for} \quad \nu = 0 \ldots 7, \tag{2}$$

where $c'_\nu \in \mathbb{C}$, $e_0 = 1$, and $e_i e_j = -\delta_{ij} + f_{ijk} e_k$ for $i, j, k \in \{1, 2, \ldots 7\}$. Here, $f_{ijk}$ is a totally anti-symmetric tensor with $f_{ijk} = 1$ when $ijk \in \{124, 235, 346, 457, 561, 672, 713\}$. The remaining values of $f_{ijk}$ are determined by anti-symmetry, and vanish otherwise. Again, the complex imaginary unit $i$ commutes with the $e_\nu$.

We may now define an element in $\mathbb{A}$ as

$$c_{\mu\nu} \, \epsilon_\mu e_\nu \qquad \text{for} \quad \mu = 0 \ldots 3, \ \nu = 0 \ldots 7, \tag{3}$$

where $c_{\mu\nu} \in \mathbb{C}$, and $\epsilon_\mu e_\nu = e_\nu \epsilon_\mu$. $\mathbb{A}$ is a $32\mathbb{C}$-dimensional non-commutative, non-associative algebra.

*2.2. Left and right multiplication*

Readers should note that the left- and right-multiplicative actions of elements in $\mathbb{A}$ can have non-trivial relationships.

For $r \in \mathbb{R} \subset \mathbb{A}$, and $c \in \mathbb{C} \subset \mathbb{A}$, left and right multiplication are identical,

$$r\Psi = \Psi r, \qquad c\Psi = \Psi c \tag{4}$$

for any $\Psi \in \mathbb{A}$. However, for a typical $h \in \mathbb{H} \subset \mathbb{A}$,

$$h\Psi \neq \Psi h. \tag{5}$$

In fact, in typical cases, it is not possible to find an $h_1 \in \mathbb{H}$ such that $h_1 \Psi = \Psi h_2$ for a given $h_2 \in \mathbb{H}$.

So far, these properties materialize as one might expect. However, what is perhaps more surprising is the fact that right multiplication by the octonions, $\mathbb{O} \subset \mathbb{A}$ *can* always be re-expressed in terms of its left multiplication algebra. For example, one finds that

$$\Psi e_7 = \tfrac{1}{2} \left( -e_7 \, \Psi + e_1 \, (e_3 \, \Psi) + e_2 \, (e_6 \, \Psi) + e_4 \, (e_5 \, \Psi) \right)$$
$$=: E_7 \Psi \tag{6}$$

for all $\Psi \in \mathbb{A}$. The left-right correspondence for the other six imaginary units may be easily found using the index cycling property

$$e_i e_j = e_k \quad \Rightarrow \quad e_{i+1} e_{j+1} = e_{k+1}. \tag{7}$$

*2.3. From $\mathbf{16} \oplus \mathbf{16}^*$ to $\mathbf{16} \oplus \mathbf{16}$*

In the standard formalism, the most common way of writing down $spin(10)$'s action on a $32\mathbb{C}$-dimensional spinor $\psi$ is as left multiplication of the bivectors $\gamma_i \gamma_j$ in $\mathbb{C}l(10)$

$$\delta \psi = r_{ij} \gamma_i \gamma_j \psi \tag{8}$$

for $i, j \in \{1, \ldots, 10\}$, $i \neq j$. It is this action which then causes the $32\mathbb{C}$-dimensional spinor to split into a problematic $\mathbf{16} \oplus \mathbf{16}^*$ pair.

Unfortunately, simply translating this scenario into the language of $\mathbb{A}$ does not resolve the problem, and so in this case, $\Psi \in \mathbb{A}$ also transforms as a **16** ⊕ **16**$^*$.

However, there is more than one way to write $spin(10)$ acting on $\mathbb{A}$. One such action was found by us a couple of years ago, and generalizes the familiar way that $spin(3,1)$ acts on spinors in $\mathbb{C} \otimes \mathbb{H}$. That is, instead of having $spin(10)$ act as in equation (8), these bivectors act in a way that generalizes the form

$$\delta \Psi_\mathrm{D} = L\Psi_\mathrm{D} P + L^* \Psi_\mathrm{D} P^*, \tag{9}$$

where $L$ represents an element of $spin(3,1)$, $P$ is a chirality projector, and $\Psi_\mathrm{D}$ is a Dirac spinor, each in $\mathbb{C} \otimes \mathbb{H}$. Here, $*$ is an automorphism of the algebra. For further detail please see the *complex invariant action* in Section 3.5 of [15]. We will not expand on this construction here, but rather in future papers.

For our current discussion we will focus now on a construction which leads to the same result by instead *generalizing the familiar Pauli matrices*. Let us define

$$\begin{aligned}
\sigma_j &:= -e_j \epsilon_2 \mid 1, & \sigma_8 &:= -i\epsilon_1 \mid 1, \\
\sigma_9 &:= -i\epsilon_3 \mid 1, & \sigma_{10} &:= -i \mid 1,
\end{aligned} \tag{10}$$

for $j \in \{1 \ldots 7\}$. Here an operator $a \mid b$ is understood to act on $\Psi \in \mathbb{A}$ as $a\Psi b$ for $a, b \in \mathbb{A}$.

With these generalized Pauli matrices, the 45 generators of $so(10)$ acting on $\Psi$ may be expressed simply as

$$\frac{1}{2} \sigma_{[a} \bar{\sigma}_{b]} \Psi = \frac{1}{4} \left( \sigma_a (\bar{\sigma}_b \Psi) - \sigma_b (\bar{\sigma}_a \Psi) \right), \tag{11}$$

for $a, b \in \{1, \ldots, 10\}$, with $a \neq b$. We define $\bar{\sigma}_c := -\sigma_c$ for $c \in \{1, \ldots, 9\}$, while $\bar{\sigma}_{10} := \sigma_{10}$. Readers may recognize that this action treats $\mathbb{A}$ as a Weyl representation. Closely related division algebraic constructions may be found in Chapter 4 of [16].

When the generalized Pauli matrices (10) are substituted into (11), we find an $so(10)$ Lie algebra which is symmetric in its treatment of quaternionic and octonionic imaginaries,

$$\delta \Psi = r_{ij}\, e_i(e_j \Psi) + r'_{mn}\, \epsilon_m(\epsilon_n \Psi) + r''_{mj}\, i\epsilon_m e_j \Psi \tag{12}$$

for $r_{ij}, r'_{mn}, r''_{mj} \in \mathbb{R}$, $i, j \in \{1, \ldots, 7\}$, $m, n \in \{1, 2, 3\}$, and $\Psi \in \mathbb{A}$.

Given equations (10)-(12), readers may confirm directly that these operators do indeed give a representation of $spin(10)$, and furthermore, that they cause $\Psi$ to now transform as

$$\Psi \;\mapsto\; \mathbf{16} \oplus \mathbf{16}. \tag{13}$$

This finally *resolves the fermion doubling problem*.

*2.4. Spacetime symmetries and the maximal commuting subalgebra*

As a final point, we mention a feature of this model which was not anticipated.

Consider the canonical $spin(10)$ action given by equation (8). It is well known that the maximal subalgebra of $\mathbb{C}l(10) = End_\mathbb{C}(\mathbb{A})$ commuting with these $spin(10)$ generators is spanned by $\{1, \gamma_{11}\}$. That is, the trivial solution and $\gamma_{11} = \gamma_1 \gamma_2 \cdots \gamma_{10}$. In other words, we find that the canonical action (8) does not leave enough freedom within $\mathbb{C}l(10)$ so as to also accommodate $sl(2, \mathbb{C})$ spacetime generators.

On the other hand, when we search for the maximal subalgebra commuting with our new $spin(10)$ generators of equation (11), we find a space spanned by $\{1, \; 1 \mid \epsilon_j\}$ over $\mathbb{C}$, with $j \in \{1, 2, 3\}$. Of course, this is none other than the trivial solution plus $sl(2, \mathbb{C})$. Furthermore, readers may confirm that this $sl(2, \mathbb{C})$ is *precisely that which is needed in order to describe Lorentz transformations on our fermions*.

While all of the internal gauge symmetries in this model can be written exclusively using left multiplication on $\Psi$, these $sl(2, \mathbb{C})$ transformations are written exclusively using right multiplication. Therefore, this model can separate internal gauge symmetries and $sl(2, \mathbb{C})$ spacetime symmetries cleanly. In fact, given that this $sl(2, \mathbb{C})$ commutes with $spin(10)$, the model is guaranteed never to run foul of the Coleman-Mandula theorem.

## 3. Standard model gauge symmetries

For the remainder of this article, we will now detail the standard model gauge symmetries, its gauge bosons, Higgs boson, and a generation of fermions in the compact language of $\mathbb{R} \otimes \mathbb{C} \otimes \mathbb{H} \otimes \mathbb{O}$.

We begin with the standard model gauge symmetries, which will be embedded inside $so(10)$ in the usual way. With the goal in mind of keeping this presentation as transparent as possible, we write these generators out explicitly in a particular basis. In this basis, $su(3)_\mathrm{C}$ appears as the subalgebra of octonionic derivations $g_2$ which holds the octonionic imaginary unit $e_7$ fixed.

A generic element of the standard model internal symmetry Lie algebra will be written as

$$\vec{g}_\mathrm{sm} \;:=\; \vec{g}_{su(3)_\mathrm{C}} + \vec{g}_{su(2)_\mathrm{L}} + \vec{g}_{u(1)_\mathrm{Y}}. \tag{14}$$

We then represent colour symmetry as

$$\vec{g}_{su(3)_\mathrm{C}} := \sum_{j=1}^{8} r_j\, i\Lambda_j \mid 1, \tag{15}$$

with $r_j \in \mathbb{R}$, and

$$\begin{aligned}
i\Lambda_1 &:= \tfrac{1}{2}(e_{34} - e_{15}) & i\Lambda_2 &:= \tfrac{1}{2}(e_{14} + e_{35}) \\
i\Lambda_3 &:= \tfrac{1}{2}(e_{13} - e_{45}) & i\Lambda_4 &:= -\tfrac{1}{2}(e_{25} + e_{46}) \\
i\Lambda_5 &:= \tfrac{1}{2}(e_{24} - e_{56}) & i\Lambda_6 &:= -\tfrac{1}{2}(e_{16} + e_{23}) \\
i\Lambda_7 &:= -\tfrac{1}{2}(e_{12} + e_{36}) & i\Lambda_8 &:= -\tfrac{1}{2\sqrt{3}}(e_{13} + e_{45} - 2e_{26}).
\end{aligned} \tag{16}$$

The operators $e_{ab}$ are understood to act on $\Psi \in \mathbb{A}$ as $e_a(e_b \Psi)$. Readers will notice that colour symmetry is written using the purely octonionic part of $\mathbb{A}$.

We subsequently find weak isospin to be represented as

$$\vec{g}_{su(2)_\mathrm{L}} := \sum_{k=9}^{11} r_k\, i\tau_k \mid 1, \tag{17}$$

where $r_k \in \mathbb{R}$,

$$\tau_9 := \tfrac{i}{2} s \epsilon_1 \qquad \tau_{10} := \tfrac{i}{2} s \epsilon_2 \qquad \tau_{11} := \tfrac{i}{2} s \epsilon_3, \tag{18}$$

and $s$ is defined as $s := \tfrac{1}{2}(1 + ie_7)$. Finally, our twelfth symmetry degree of freedom, weak hypercharge, is given by

$$\vec{g}_{u(1)_\mathrm{Y}} := r_{12}\, iY \mid 1, \tag{19}$$

where $r_{12} \in \mathbb{R}$ and

$$Y := -\frac{i}{6}(e_{13} + e_{26} + e_{45}) + \frac{i}{2} s^* \epsilon_3. \tag{20}$$

For completion, we mention that a $B - L$ symmetry may additionally be represented as

$$\vec{g}_{u(1)_{\text{B-L}}} \;:=\; r_0\, i\,(B - L) \mid 1, \tag{21}$$

where $r_0 \in \mathbb{R}$, and

$$B - L := -\frac{i}{3}(e_{13} + e_{26} + e_{45}). \quad (22)$$

With this purely division algebraic construction of the standard model's gauge symmetries, one might now wonder how these symmetries behave under a familiar automorphism of the algebra: complex conjugation.

## 4. Unbroken gauge symmetries

Readers are now encouraged to verify for themselves that the condition

$$\vec{g}_{sm}{}^* = \vec{g}_{sm} \quad (23)$$

is satisfied by only a subset of the standard model's internal symmetry generators. This subalgebra happens to be none other than

$$\vec{g}_{ub} := \vec{g}_{su(3)_C} + \vec{g}_{u(1)_{EM}}, \quad (24)$$

the *standard model's known unbroken gauge symmetries*.

Explicitly, $\vec{g}_{su(3)_C}$ is described as in equation (16), and $\vec{g}_{u(1)_{EM}}$ is given by

$$\vec{g}_{u(1)_{EM}} := r_{13} \, iQ \mid 1, \quad (25)$$

where

$$Q := -\frac{i}{6}(e_{13} + e_{26} + e_{45}) + \frac{i}{2}\epsilon_3. \quad (26)$$

To date, the Higgs mechanism has been confirmed experimentally as a means to break $\vec{g}_{sm}$ to $\vec{g}_{ub}$. However, from a theoretical perspective, it is unclear as to why Nature should choose *this* Higgs so as to break $\vec{g}_{sm}$ in this particular way. *Why not some other subalgebra?* Clearly, it is worth investigating whether this algebraic characterization can lead to an answer.

## 5. $SL(2, \mathbb{C})$ spacetime symmetries

We may now describe $sl(2, \mathbb{C})$ spacetime symmetries, $\vec{g}_{st}$, via right multiplication by a 6 $\mathbb{R}$-dimensional subspace of $\mathbb{C} \otimes \mathbb{H}$ on $\mathbb{A}$. Explicitly,

$$\vec{g}_{st} := 1 \mid c_j \epsilon_j \quad (27)$$

for $j \in \{1, 2, 3\}$, and $c_j \in \mathbb{C}$.

## 6. Fermion identification

With the action of $\vec{g}_{sm}$, $\vec{g}_{u(1)_{B-L}}$, and $\vec{g}_{st}$ defined on $\Psi \in \mathbb{A}$, we are now in a position to label the degrees of freedom in $\Psi$.

Let us define a new basis for $\mathbb{C} \otimes \mathbb{H}$ as

$$\begin{aligned}
\epsilon_{\uparrow\uparrow} &:= \tfrac{1}{2}(1 + i\epsilon_3) & \epsilon_{\downarrow\downarrow} &:= \tfrac{1}{2}(1 - i\epsilon_3) \\
\epsilon_{\uparrow\downarrow} &:= \tfrac{1}{2}(-\epsilon_2 + i\epsilon_1) & \epsilon_{\downarrow\uparrow} &:= \tfrac{1}{2}(\epsilon_2 + i\epsilon_1).
\end{aligned} \quad (28)$$

Similarly, we define a new basis for $\mathbb{C} \otimes \mathbb{O}$ as

$$\begin{aligned}
\ell &:= \tfrac{1}{2}(1 + ie_7) & \ell^* &:= \tfrac{1}{2}(1 - ie_7) \\
q_1 &:= \tfrac{1}{2}(-e_5 + ie_4) & q_1^* &:= -\tfrac{1}{2}(e_5 + ie_4) \\
q_2 &:= \tfrac{1}{2}(-e_3 + ie_1) & q_2^* &:= -\tfrac{1}{2}(e_3 + ie_1) \\
q_3 &:= \tfrac{1}{2}(-e_6 + ie_2) & q_3^* &:= -\tfrac{1}{2}(e_6 + ie_2).
\end{aligned} \quad (29)$$

Extracting $\vec{g}_{sm}$, $\vec{g}_{u(1)_{B-L}}$, and $\vec{g}_{st}$'s combined Cartan subalgebra now supplies us with enough such operators to fully identify our fermionic states.

$$\begin{aligned}
\Psi =\ & \left(\mathcal{V}_L^\uparrow \epsilon_{\uparrow\uparrow} + \mathcal{V}_L^\downarrow \epsilon_{\uparrow\downarrow} + \mathcal{E}_L^{-\uparrow} \epsilon_{\downarrow\uparrow} + \mathcal{E}_L^{-\downarrow} \epsilon_{\downarrow\downarrow}\right)\ell \\
& + \left(\mathcal{E}_R^{-\downarrow *} \epsilon_{\uparrow\uparrow} - \mathcal{E}_R^{-\uparrow *} \epsilon_{\uparrow\downarrow} - \mathcal{V}_R^{\downarrow *} \epsilon_{\downarrow\uparrow} + \mathcal{V}_R^{\uparrow *} \epsilon_{\downarrow\downarrow}\right)\ell^* \\
& - i\left(\mathcal{U}_L^{a\uparrow} \epsilon_{\uparrow\uparrow} + \mathcal{U}_L^{a\downarrow} \epsilon_{\uparrow\downarrow} + \mathcal{D}_L^{a\uparrow} \epsilon_{\downarrow\uparrow} + \mathcal{D}_L^{a\downarrow} \epsilon_{\downarrow\downarrow}\right) q_a \\
& + i\left(\mathcal{D}_R^{a\downarrow *} \epsilon_{\uparrow\uparrow} - \mathcal{D}_R^{a\uparrow *} \epsilon_{\uparrow\downarrow} - \mathcal{U}_R^{a\downarrow *} \epsilon_{\downarrow\uparrow} + \mathcal{U}_R^{a\uparrow *} \epsilon_{\downarrow\downarrow}\right) q_a^*,
\end{aligned} \quad (30)$$

where the complex coefficients, $\mathcal{V}_L^\uparrow$, $\mathcal{V}_L^\downarrow$, ... are labelled in an obvious way. For example, $\mathcal{V}_L^\uparrow$, represents a left-handed neutrino, where the upward-pointing arrow indicates a $+1/2$ eigenvalue under the $\tfrac{1}{2}i\epsilon_3 = \tfrac{1}{2}\sigma_z$ operator.

Readers may see that $\Psi$ now represents one generation in its left-handed Weyl representation form. It should be noted, of course, that it still includes those singlet states under $\vec{g}_{su(2)_L}$, associated with right-handed particles.

It is of special interest to note a further richness we see in this algebraic formulation of standard model states. Namely, readers may appreciate that $\ell$ and $\ell^*$, related to leptons, are idempotents, $\ell\ell = \ell$, and $\ell^*\ell^* = \ell^*$. On the other hand, $q_a$ and $q_a^*$, related to quarks, are nilpotent, $q_a q_a = 0$, and $q_a^* q_a^* = 0$, with no implied sum on $a$. We propose revisiting the possibility that such properties could ultimately relate to quark confinement, [5].

## 7. Gauge and Higgs bosons

Readers may not be surprised to learn that derivatives, even covariant ones, can easily be written in a division algebraic language. For example,

$$\partial_\mu \Psi \, \hat{\sigma}^\mu := \partial_0 \Psi - \partial_1 \Psi i\epsilon_1 + \partial_2 \Psi i\epsilon_2 - \partial_3 \Psi i\epsilon_3, \quad (31)$$

where the minus signs here appear as a consequence of our use of *right* multiplication by the Pauli operators.

Furthermore, it is straightforward to confirm that the full covariant derivative extends equation (31) naturally as

$$D\Psi :=$$
$$\left(\partial_\mu \Psi - \tfrac{i}{2} g_3 G_\mu^m \Lambda_m \Psi - \tfrac{i}{2} g_2 W_\mu^i \tau_i \Psi - \tfrac{i}{2} g_1 B_\mu Y \Psi\right)\hat{\sigma}^\mu, \quad (32)$$

for coupling constants $g_1, g_2, g_3 \in \mathbb{R}$, and $G_\mu^m, W_\mu^i, B_\mu \in \mathbb{R}$.

Finally, we write the familiar standard model Higgs as a *pure quaternion*

$$\begin{aligned}
h &:= \phi^{0*} \epsilon_{\uparrow\uparrow} - \phi^{+*} \epsilon_{\downarrow\uparrow} + \phi^+ \epsilon_{\uparrow\downarrow} + \phi^0 \epsilon_{\downarrow\downarrow} \\
&= Re(\phi^0)\epsilon_0 - Im(\phi^+)\epsilon_1 \\
& \quad -Re(\phi^+)\epsilon_2 + Im(\phi^0)\epsilon_3 \quad \in \mathbb{H},
\end{aligned} \quad (33)$$

for $\phi^0$ and $\phi^+$ in $\mathbb{C}$. We furthermore define a coupling operator $\hat{k}$ as

$$\hat{k} := -8\left(f \epsilon_{\uparrow\uparrow} + b \epsilon_{\downarrow\downarrow}\right) s^* S^* - 8\left(d \epsilon_{\uparrow\uparrow} + g \epsilon_{\downarrow\downarrow}\right) s^* S, \quad (34)$$

with $f, b, d, g \in \mathbb{R}$, and $S := \tfrac{1}{2}(1 + iE_7)$ as per equation (6). We define the product of $h$ and $\hat{k}$ as $H := h\hat{k}$.

In an upcoming companion paper, [2], we show explicitly, for the first time we believe, the phenomenon of *quaternionic triality* within a left-right symmetric Higgs system.

## 8. Scalars

With these particle representations now specified, we may begin writing scalars invariant under $\vec{g}_{sm}$, $\vec{g}_{u(1)_{B-L}}$, and $\vec{g}_{st}$. It so happens that certain kinetic and Yukawa scalar invariants from the standard model find here a succinct form:

$$\langle \Psi^{\dagger}(D\Psi) \rangle \quad \text{and} \quad \langle \widetilde{\Psi}(H\Psi) \rangle, \tag{35}$$

up to overall constant factors. The brackets $\langle \dots \rangle$ here mean to take the real part.

The operations $\dagger$ and $\sim$ each symbolize an involution on $\mathbb{A}$. Specifically, the involution $\sim$ maps $e_j \mapsto -e_j$ for $j \in \{1, \dots, 7\}$, and $\epsilon_k \mapsto -\epsilon_k$ for $k \in \{1, 2, 3\}$, while reversing the order of multiplication:

$$\widetilde{a_1 a_2} = \widetilde{a}_2 \widetilde{a}_1 \tag{36}$$

for $a_1, a_2 \in \mathbb{A}$. We define the involution $\dagger$ to be the composition of the involution $\sim$ together with complex conjugation, $*$.

## 9. Conclusion

In this article, we solve the $\mathbb{R} \otimes \mathbb{C} \otimes \mathbb{H} \otimes \mathbb{O}$ fermion doubling problem. That is, we demonstrate for the first time a generalization of Günaydin and Gürsey's early colour quark triplet, where one full generation is now identified with just one copy of $\mathbb{R} \otimes \mathbb{C} \otimes \mathbb{H} \otimes \mathbb{O}$,

$$\mathbb{R} \otimes \mathbb{C} \otimes \mathbb{H} \otimes \mathbb{O} \quad \leftrightarrow \quad \text{one generation.}$$

Unlike with previous attempts, this does not involve a doubling of the algebra, nor a proliferation into matrices and column vectors. Our new description compiles one generation of fermionic states into the form of a left-handed Weyl representation. This result should translate easily so as to solve the analogous problem faced by well-known $\mathbb{C}l(10)$ models involving $32\mathbb{C}$-dimensional spinors.

Finally, we restrict our standard model internal symmetries, $\vec{g}_{sm}$, to be invariant under the familiar complex conjugate, $i \mapsto -i$. This results in sending $\vec{g}_{sm}$ to the standard model's two unbroken gauge symmetries,

$$\vec{g}_{sm} \quad \mapsto \quad \vec{g}_{su(3)_C} + \vec{g}_{u(1)_{EM}}. \tag{37}$$

Clearly, it is worth investigating whether or not this algebraic feature ultimately shepherds the standard model's collapse to $su(3)_C \oplus u(1)_{EM}$, as opposed to any other subalgebra.

## Declaration of competing interest

The authors declare that they have no known competing financial interests or personal relationships that could have appeared to influence the work reported in this paper.

## Acknowledgements

The results of this paper were first presented in a recorded talk for Rutgers University Department of Mathematics, 29th of October, 2020, and subsequently in a recorded talk for Nikhef, 5th of November, 2020. An early version of the paper was circulated widely amongst colleagues on the 16th of February, 2021. Subsequently, it was presented at Perimeter Institute on the 22nd of February, 2021, http://pirsa.org/21020027/, again at Perimeter Institute, http://pirsa.org/21030013/, and most recently at the University of Edinburgh, empg.maths.ed.ac.uk/HTML/AY2020Seminars.html

This work was graciously supported by the VW Stiftung Freigeist Fellowship and a visiting fellowship at the African Institute for Mathematical Sciences in Cape Town. Furthermore, we are grateful for questions and feedback on this work from Michael Borinsky, Latham Boyle, David Chester, Michael Duff, Sheldon Goldstein, Brage Gording, Judd Harrison, Franz Herzog, John Huerta, Alessio Marrani, Piet Mulders, Agostino Patella, Jan Plefka, Mike Rios, Beth Romano, Matthias Staudacher, Shadi Tahvildar-Zadeh, Ivan Todorov, Stijn van Tongeren, and others we may have inexcusably missed.

## References

[1] C. Furey, $SU(3)_C \times SU(2)_L \times U(1)_Y (\times U(1)_X)$ as a symmetry of division algebraic ladder operators, Eur. Phys. J. C 78 (5) (2018) 375.
[2] N. Furey, M.J. Hughes, Division algebraic symmetry breaking, in preparation, https://pirsa.org/21030013.
[3] A. Conway, Quaternion treatment of relativistic wave equation, Proc. R. Soc. Lond. Ser. A, Math. Phys. Sci. 162 (909) (1937).
[4] M. Günaydin, F. Gürsey, Quark structure and the octonions, J. Math. Phys. 14 (11) (1973).
[5] M. Günaydin, Octonionic Hilbert spaces, the Poincare group and SU(3), J. Math. Phys. 17 (1976) 1875, https://doi.org/10.1063/1.522811.
[6] A. Barducci, F. Buccella, R. Casalbuoni, L. Lusanna, E. Sorace, Quantized Grassmann variables and unified theories, Phys. Lett. B 67 (1977) 344.
[7] R. Casalbuoni, R. Gatto, Unified description of quarks and leptons, Phys. Lett. B 88 (1979) 306.
[8] G. Dixon, Division Algebras: Octonions Quaternions Complex Numbers and the Algebraic Design of Physics, Kluwer Academic Publishers, 1994.
[9] Z.G. Silagadze, SO(8) colour as possible origin of generations, Phys. At. Nucl. 58 (1995) 1430–1434; Yad. Fiz. 58 (8) (1995) 1513–1517, arXiv:hep-ph/9411381.
[10] C.A. Manogue, T. Dray, Octonions, E6, and particle physics, J. Phys. Conf. Ser. 254 (2010) 012005.
[11] G. Dixon, Division algebras; spinors; idempotents; the algebraic structure of reality, arXiv:1012.1304v1 [hep-th].
[12] A. Anastasiou, L. Borsten, M.J. Duff, M.J. Hughes, S. Nagy, Super Yang-Mills, division algebras and triality, J. High Energy Phys. 1408 (2014) 080, arXiv:1309.0546.
[13] C. Furey, Generations: three prints, in colour, J. High Energy Phys. 10 (2014) 046, arXiv:1405.4601 [hep-th].
[14] C. Furey, Charge quantization from a number operator, Phys. Lett. B 742 (2015) 195–199, arXiv:1603.04078 [hep-th].
[15] C. Furey, Standard model physics from an algebra?, PhD thesis, University of Waterloo, 2015, www.repository.cam.ac.uk/handle/1810/254719, arXiv:1611.09182 [hep-th].
[16] M.J. Hughes, Octonions and Supergravity, PhD thesis, Imperial College London, 2016, https://spiral.imperial.ac.uk/handle/10044/1/34938.
[17] C. Burdik, S. Catto, Y. Gürcan, A. Khalfan, L. Kurt, Revisiting the role of octonions in hadronic physics, Phys. Part. Nucl. Lett. 14 (2) (2017) 390–394.
[18] P. Bolokhov, Quaternionic wavefunction, IJMPA 34 (2019) 1950001, arXiv:1712.04795 [quant-ph].
[19] N. Gresnigt, Braids, normed division algebras, and standard model symmetries, Phys. Lett. B 783 (2018).
[20] C. Furey, Three generations, two unbroken gauge symmetries, and one eight-dimensional algebra, Phys. Lett. B 785 (2018) 84–89, arXiv:1910.08395.
[21] I. Todorov, M. Dubois-Violette, Deducing the symmetry of the standard model from the automorphism and structure groups of the exceptional Jordan algebra, Int. J. Mod. Phys. A 33 (20) (2018) 1850118.
[22] B. Gording, A. Schmidt-May, The unified standard model, arXiv:1909.05641.
[23] C. Castro Perelman, RCHO-valued gravity as a grand unified field theory, Adv. Appl. Clifford Algebras 29 (1) (2019) 22.
[24] C. Castro Perelman, On CHO-valued gravity, sedenions, hermitian matrix geometry and nonsymmetric Kaluza-Klein theory, Adv. Appl. Clifford Algebras 29 (3) (2019) 58.
[25] K. Krasnov, SO(9) characterisation of the standard model gauge group, arXiv:1912.11282.
[26] D. Chester, M. Rios, A. Marrani, Beyond the standard model with six-dimensional spacetime, arXiv:2002.02391.
[27] L. Boyle, The standard model, the exceptional Jordan algebra, and triality, arXiv:2006.16265.
[28] T. Asselmeyer-Maluga, Braids, 3-manifolds, elementary particles: number theory and symmetry in particle physics, Symmetry 11 (10) (2019) 1298.
[29] T. Singh, Trace dynamics and division algebras: towards quantum gravity and unification, Z. Naturforsch. A 76 (2021) 131, arXiv:2009.05574 [hep-th].
[30] I. Todorov, Superselection of the weak hypercharge and the algebra of the standard model, arXiv:2010.15621.
[31] J.M. Gracia-Bondia, B. Iochum, T. Schucker, The standard model in noncommutative geometry and fermion doubling, Phys. Lett. B 416 (1998) 123, arXiv:hep-ph/9709145.
[32] A. Bochniak, A. Sitarz, A spectral geometry for the standard model without fermion doubling, arXiv:2001.02902 [hep-th].
[33] M. Dubois-Violette, I. Todorov, Superconnection in the spinfactor approach to particle physics, Nucl. Phys. B 957 (2020) 115065, arXiv:2003.06591 [hep-th].